\documentstyle[preprint,aps]{revtex}
\newcommand{\be}{\begin{eqnarray}}
\newcommand{\ee}{\end{eqnarray}}
\newcommand{\ra}{\rightarrow}
\newcounter{appendixc}
\newcounter{subappendixc}[appendixc]
\newcounter{subsubappendixc}[subappendixc]

\renewcommand{\appendix}[1] {\vspace{0.6cm}
        \refstepcounter{appendixc}
        \setcounter{table}{0}
        \setcounter{equation}{0}
        \renewcommand{\thetable}{\Alph{appendixc}.\arabic{table}}
        \renewcommand{\theappendixc}{\Alph{appendixc}}
        \renewcommand{\theequation}{\Alph{appendixc}.\arabic{equation}}
        \noindent{\bf Appendix \theappendixc #1}\par\vspace{0.4cm}}

\begin{document}
\draft
\preprint{KAIST--CHEP--94/13}
%\hspace{-38.0mm}
%\raisebox{2.4ex}{}}
%\preprint{SNUTP--94--XX}
\title{$B \rightarrow {\tau}{\nu}X$ decays in a two Higgs doublet model  }
\author{   
Yeong Gyun Kim\thanks{ygkim@chep6.kaist.ac.kr}, 
Jae Yong Lee\thanks{jylee@chep5.kaist.ac.kr}, 
Kang Young Lee\thanks{kylee@chep5.kaist.ac.kr} 
and Jae Kwan Kim
}
\vspace{.6in}
\address{           
%$^{b}$ , KOREA  \\ 
Dept. of Physics, KAIST, Taejon 305-701, KOREA
}
\date{\today}
\maketitle
\begin{abstract}
\\
We calculate the decay rate for inclusive 
$B \rightarrow {\tau}{\nu}X$ decays in a two Higgs doublet model 
using heavy quark expansion and operator product expansion. 
Combined with the recent measurement of Br($B \ra \tau \nu X$), 
we find a limit $\tan \beta < 0.58 m_H/\mbox{GeV}$ at 90\%\ C.L..
\end{abstract}
\pacs{ }
% body of paper here

\narrowtext
%\tighten

\label{sec:intro}

One of the most exciting topic in the Standard Model (SM) is
the existence of the Higgs boson and its nature.
Even though there is only one neutral Higgs boson in the
SM, more complicated structure of the Higgs boson sector
is usually assumed in many models beyond the SM \cite{HH}.
The two Higgs doublet model is one of the minimal extensions of the SM.
One or more charged Higgs bosons inevitably exist 
when the multi-Higgs doublets are assumed. 
Hence it is interesting to find any of the charged Higgs boson effects 
in probing new physics.
With the CLEO bound, the flavour-changing process 
$b \ra s \gamma$ is known to provide the most
stringent constraint on the charged Higgs parameter space \cite{CLEO}.
It is also reported recently that the vertex $Z \ra b \bar{b}$
with one - loop corrections can also provide additional
constraints for $\tan \beta \lesssim 1$ \cite{park}.

The inclusive semileptonic decays of B meson involving
a tau lepton in the final state can be also an useful probe for
the charged Higgs effects of multi-Higgs models.
This process is one of the latest measured channel which
constitutes a significant fraction of all semileptonic B decays.
The ALEPH and L3 collaborations at LEP $e^+e^-$ collider 
reported on the improved measurement of branching ratios of the decay
$\bar{B} \ra \tau \bar{\nu} X$ \cite{putzer,l3}:
\be
\mbox{Br}(\bar{B} \ra X \tau \bar{\nu}) &=& 
       (2.76 \pm 0.47 (stat.) \pm 0.43 (syst.)) \%~~~~~~~~~~~\mbox{ALEPH}
	\nonumber \\
\mbox{Br}(\bar{B} \ra X \tau \bar{\nu}) &=& 
       (2.4 \pm 0.7 (stat.) \pm 0.8 (syst.)) \%~~~~~~~~~~~~~~~~~~\mbox{L3}
	\nonumber 
\ee
Being distinct from other probes of charged Higgs effects, this process
is not sensitive to the top quark mass. Furthermore we can do the
systematic study for the nonperturbative contributions of this decay
process in the framework of heavy quark effective theory(HQET).

Recently, there has been a great progress in systematic study of
inclusive semileptonic decays of hadrons containing a single
heavy quark. 
In the pioneering work by Chay, Georgi and Grinstein \cite{chay},
the method by means of the operator product expansion (OPE) in
the heavy quark effective theory was suggested, 
which is motivated by deep inelastic scattering.
Following this formalism, one can implement the QCD - based
expansion in powers of $1/m_Q$ to calculate inclusive
$\bar{B} \ra l \bar{\nu} X$ decay rates and lepton spectra.
The leading order terms coincides with those of the free quark decay.
The noticeable result is that the leading nonperturbative corrections 
are of order of $1/m^2_Q$ because terms of order of $1/m_Q$ vanish.
Several authors have studied the inclusive 
$\bar{B} \ra \tau \bar{\nu} X$ decay rates in the framework of the SM
\cite{falk,balk,koyrakh}.

Here we calculate the decay rates and lepton spectra for the
the inclusive semileptonic decays of B meson into a tau lepton
with the help of the HQET in a two Higgs doublet model, so-called
type II. 
In the type II model, one Higgs doublet couples to up type quark sector
and the other Higgs doublet to down type quark sector,
while only one Higgs doublet couples to the fermions in type I model.
We analyse only type II model because it is
required in the minimal supersymmetric standard model (MSSM)
and type I is less interesting as type I is similar to the SM.
We also discuss their phenomenological meaning as
constraint for the charged Higgs boson.

In the two Higgs doublet model of type II, the inclusive $\bar{B} \ra \tau \bar{\nu} X$ decay
is due to the hamiltonian density,
\be
{\cal H} &=& {\cal H}_W + {\cal H}_H
\ee
where
\be
{\cal H}_W &=& - \frac{{G_F}^2}{\sqrt{2}} V_{cb} \,\, J_{\mu} J_{\tau}^{\mu}
              \nonumber \\
{\cal H}_H &=& - \frac{{G_F}^2}{\sqrt{2} m_H^2} V_{cb} \,\, 
                 \tilde{J} \tilde{J}_{\tau}     
\ee
with
\be
J_{\mu} &=& \bar{c} \gamma_\mu (1-\gamma_5) b\,\,
              \nonumber \\
J_{\tau}^{\mu} &=& \bar{\tau} \gamma^\mu (1-\gamma_5) \nu_{\tau}
              \nonumber \\
\tilde J &=& \bar{c}~[m_b X (1+\gamma_5) + m_c Y (1-\gamma_5)]~b
              \nonumber \\
{\tilde J}_\tau &=& \bar{\tau} ~[m_{\tau} X (1-\gamma_5)]~ \nu_{\tau}
\ee
The ratio of two vacuum expectation values are defined by
\be
X = \tan \beta \equiv \frac{v_2}{v_1}  \nonumber
\ee
and $Y=1/X$.
The inclusive differential decay rate for the process 
$\bar{B}(p_B) \ra X(p_X) \tau(p_\tau) \bar{\nu}(p_{\bar{\nu}})$
\be
\mbox {d}\Gamma & = &
            \frac{|V_{cb}|^2 \, {G_F}^2}{2 \pi^3 }
                  \{ \,\,L_{\mu \nu} \, W^{\mu \nu}  
 + \frac{m^2_\tau X^2}{m^4_{H^+}} (p_\tau \cdot p_{\bar{\nu}}) \,W
  - \frac{m_\tau X}{{m^2}_{H^+}}
 (\frac{1}{2} m_\tau {p_{\bar{\nu}}}_\mu)\,{W^\mu}\,\,\} 
            \nonumber \\
         &&   \times dq^2~dE_{\tau}~ dE_{\nu}
\ee
is related to the three hadronic quantaties,
\be
W^{\mu \nu}&=& \sum_X (2 \pi)^3 \delta^4 (p_B - q - p_X )
 \langle B(v) | {J^{\mu}}^\dagger | X \rangle \langle X | J^\nu | B(v) \rangle 
            \nonumber \\
W^{\mu} &=&  \sum_X (2 \pi)^3 \delta^4 (p_B - q - p_X )
 \langle B(v) | {J^\mu}^\dagger | X \rangle \langle X | \tilde J | B(v)\rangle 
          + H.c.
            \nonumber \\
W &=& \sum_X (2 \pi)^3 \delta^4 (p_B - q - p_X )
 \langle B(v) | {\tilde J}^\dagger | X \rangle 
                  \langle X | \tilde J | B(v) \rangle 
\ee
where leptonic tensor $L^{\mu \nu}$ is given by 
\be
L^{\mu \nu} = 
{p_\tau}^\mu {p_{\bar{\nu}}}^\nu + {p_\tau}^\nu {p_{\bar{\nu}}}^\mu 
- g^{\mu \nu}{p_\tau} \cdot {p_{\bar{\nu}}}
+i \epsilon^{\mu \nu \rho \sigma } {p_\tau}_\rho {p_{\bar{\nu}}}_\sigma
\ee
and $q = p_\tau+p_{\bar{\nu}}$ is the momentum transfer.
We can expand the quantities in terms of form factors.
The hadronic tensor $W^{\mu \nu}$ is parameterized by
\begin{equation}
W^{\mu \nu}=
-g^{\mu\nu} W_1 + v^{\mu}v^{\nu} W_2 - i
\epsilon^{\mu\nu\alpha\beta} v_\alpha q_\beta W_3 + q^\mu q^\nu W_4 +
(q^\mu v^\nu + q^\nu v^\mu) W_5\,\,,
\end{equation}
and
\begin{equation}
{W^\mu}=q^\mu V_1 + v^\mu V_2 \,\,,
\end{equation}
and
\begin{equation} 
W=S. 
\end{equation}
where $v$ is the four velocity of B meson.

Now our goal is to calculate the quantities $W^{\mu \nu}$, $W^{\mu}$, $W$ .
The form factors in $W^{\mu \nu}$, $W^{\mu}$, $W$ are determined 
by following amplitudes $T^{\mu \nu}$, $T^{\mu}$, $T$ respectively,
\be
T^{\mu \nu} &=&
-i\int d^4 x  e^{-i q \cdot x}
 \langle B(v) |T \{ {J^\mu}^\dagger(x) J^\nu(0) \} | B(v) \rangle
 \nonumber \\
T^\mu &=&
-i\int d^4 x  e^{-i q \cdot x}
 \langle B(v) |T \{ {J^\mu}^\dagger(x) \tilde J(0) + H.c. \} | B(v) \rangle
 \nonumber \\
T &=&
-i\int d^4 x  e^{-i q \cdot x}
 \langle B(v) |T \{ {\tilde J}^\dagger(x) \tilde J(0) \} | B(v) \rangle.
\ee
  The time ordered products of two currents
\be
-i \int d^4 x  e^{-i q \cdot x} T \{{ J J }\}
\ee
can be computed using an OPE in terms of operators involving b-quark fields
in the HQET. At tree level, one may obtain OPE up to operators of 
dimension five by considering the relevant diagrams in fig.1.
The calculation is straightforward and discribed in detail elsewhere
\cite{chay,bigi,manohar,blok}.
Here we just present the results.
Introducing a set of dimensionless variables 
\begin{equation}
\hat{q}^2 = \frac{q^2}{{m_b}^2}\,\,,\,\,~~~
y=\frac{2 E_\tau}{m_b}\,\,,~~~\,\,
x=\frac{2 E_\nu}{m_b}\,\,,
\end{equation}
and defining the mass ratios
\begin{equation}
\rho_c = \frac{{m_c}^2}{{m_b}^2},~~~~~~~
\rho_\tau= \frac{{m_\tau}^2}{{m_b}^2}\,\,,\,\,
\end{equation}
where $E_\tau$ and $E_\nu$ are the energies of tau lepton and neutrino in
B rest frame,the triple differential decay rate can be written by
\be
\frac{1}{\Gamma_b} \frac{d \Gamma}{d \hat{q}^2 d y d x}
&=&   24\,\Theta\left(x-\frac{2(\hat{q}^2-\rho_\tau)}{y+
      \sqrt{y^2-4 \rho_\tau}}\right)~
      \Theta\left(\frac{2(\hat{q}^2-\rho_\tau)}{y+
      \sqrt{y^2-4 \rho_\tau}}-x\right) \nonumber \\
& &\{\,\,(\hat{q}^2-{\rho_\tau}) \hat{W_1}+ \frac{1}{2}(yx-\hat{q}^2
              +  {\rho_\tau})\hat {W_2} \nonumber \\
& &+ \frac{1}{2}[\hat{q}^2 (y-x)-{\rho_\tau}(y+x)]
     \hat W_3+
     \frac{1}{2} \rho_\tau (\hat{q}^2-\rho_\tau)\hat W_4+
     \rho_\tau x \hat W_5 \nonumber \\
& & -\frac{1}{2}R~\sqrt{\rho_{\tau}}~[(\hat q^2-\rho_{\tau}) \hat V_1 
    + x \hat V_2] \nonumber \\
& & +\frac{1}{2} R^2 (\hat q^2-\rho_{\tau}) \hat S \,\,\} 
\ee
where  
\be
\Gamma_b&=& \frac{|V_{cb}|^2 {G_F}^2 {m_b}^5}{192 \pi^3}\,\,,\nonumber \\
R&=&\frac{m_b m_\tau}{m_H^2} X^2 \,\,.
\ee
The expressions for the form factors $\hat W_i$, $\hat V_i$, and
$\hat S$ at tree level and to
order $1/m_b^2$ in the operator product expansion are given 
in the appendix A.
We assumed $X > 1$ and neglected the terms propertional to $Y^2 \rho_c$.
The leading term in $1/m_b$ expansion reproduces 
the parton model results.
The nonperturbative corrections of order $1/m_b^2$ is written 
in terms of two hadronic parameters $\lambda_1$, $\lambda_2$. 
The parameters are related to the kinetic energy $K_b$
of the b-quark inside the B-meson and to the mass splitting 
between B and B$^*$ mesons.

Integrating over the kinematic variable $\hat q^2$ and $x$, 
we obtain the differential decay rates with respect to the rescaled 
lepton energy $y$ which can be written as
\be
\frac{1}{\Gamma_b} \frac{d \Gamma}{d y}~=~{\cal J}_1 
- R~{\cal J}_2  + R^2{\cal J}_3
\ee
where the rather lengthy expressions of ${\cal J}_1$, ${\cal J}_2$
and ${\cal J}_3$ are given in the appendix B.

Finally, we perform the $y$-integration over the kinametic region
\begin{equation}
2\sqrt{\rho_{\tau}} \le y \le 1+\rho_{\tau}-\rho_c
\end{equation}
to obtain the total rates. For $\Gamma$, we find
\begin{equation}\label{totaldecay}
\frac{\Gamma}{\Gamma_b} = \Gamma_1 
- R~ \Gamma_2 + R^2 \Gamma_3
\end{equation}
where again the expression of $\Gamma_1$, $\Gamma_2$ and 
$\Gamma_3$ are given in the appendix C.
As can be seen easily, our result reduces to the SM result of
the ref.\cite{falk,balk,koyrakh} if $R$=0.

From the above result, the following relations holds
\be
\Gamma(\bar B \ra \tau \bar{\nu} X) = 
        \Gamma^{\mbox{{\tiny(SM)}}}(\bar B \ra \tau \bar{\nu} X)  
        ( 1 - A R + B R^2 )
\ee
where $A = \Gamma_2/\Gamma_1$ and $B = \Gamma_3/\Gamma_1$.
Using this relation, one can obtain the constraint on $\tan \beta / m_H$ from
the measurement of branching ratio, Br($B \ra \tau \nu X$ ).
To obtain the numerical value, we need the values of the tau mass $m_\tau$, 
the heavy quark masses $m_b$ and $m_c$, the hadronic parameters 
$\lambda_1$ and $\lambda_2$ and CKM matrix element $|V_{cb}|$ 
as input parameters.
From the heavy quark expansion of B and D meson masses,
four parameters $m_b,~m_c,~\lambda_1,~\lambda_2$ are reduced 
to three parameters $\bar \Lambda,~ \lambda_1,~\lambda_2$,where
$\bar \Lambda$ is associated with the effective mass of 
light degrees of freedom in heavy mesons.
We then take the values of them used in the paper of Falk et al.\cite{falk}.
To avoid the uncertainty in CKM matrix element $|V_{cb}|$,
we normalize the branching ratio of $B \ra \tau \nu X$ 
to that of $B \ra e \nu X$ \cite{pdg}:
\be
\mbox{Br}(B \ra e \nu X) = (10.7 \pm 0.5) \%.
\ee
including the perturbative QCD correction.

This procedure leads to the following values of $A$ and $B$:
\be
A=0.43,~~~~~~~B=0.27
\ee
With the measured branching ratio of L3 \cite{l3}, this gives the refined
constraint
\be
X < \frac{0.58m_H}{\mbox{GeV}}
\ee 
at 90$\%$ C.L..

Isidori \cite{isidori} has estimated this limit with free quark
decay formulation and the first measurement of this channel in ALEPH
\cite{aleph}. The value obtained by them was $ X < 0.54 m_H/$GeV.
But they made a mistake in numerical calculation of $A$ in eq. (25).
With correcting the value of $A$, their result should be modified as 
$ X < 0.61 m_H/$GeV.  
The difference of values between their result and our one
is principally caused by the fact that we used the recent data to 
estimate our limit.
And it is also due to the differences of values of input parameters
as well as $1/m_b$ corrections.

In conclusion, we calculate the decay rate of inclusive 
$B \ra \tau \nu X$ decayin two Higgs doublet model using HQET and OPE. 
From the recent measurement of Br( $B \ra \tau \nu X$), 
we get the limit on $\tan \beta / m_H$.

\acknowledgements

This work is supported in part by Korea Science and Engineering Foundation
(KOSEF). 
\newpage
\appendix

The form factors $\hat W_i$, $\hat V_i$, and
$\hat S$ to order $1/m_b^2$ in the operator product expansion 
are given by
\be
\hat W_1 &=& \delta(\hat z) \{ \frac{1}{4}(2-y-x)
            - \frac{\lambda_1+3 \lambda_2}{12 m_b^2} \}
	     \nonumber  \\
         &&+ \delta'(\hat z) \{ \frac{\lambda_1}{24 m_b^2}~
             [ 8 \hat q^2+6 (y+x)-5 (y+x)^2 ] 
	      \nonumber \\
         &&~~~~~~+ \frac{\lambda_2}{8 m_b^2}~[8 \hat q^2 +14 (y+x) 
	   -5 (y+x)^2 -16 ] \}
	    \nonumber   \\
         &&+ \delta''(\hat z) \frac{\lambda_1}{24 m_b^2}~(2-y-x)
	    [4 \hat q^2 -(y+x)^2]\,\,,
	     \nonumber  \\
\hat W_2 &=& \delta(\hat z) \{1-\frac{5(\lambda_1+3 \lambda_2)}{6 m_b^2}
	    \} 
	     \nonumber    \\
         &&+ \delta'(\hat z) \{ \frac{7\lambda_1}{6 m_b^2}~(y+x)
	   + \frac{\lambda_2}{2 m_b^2}~[5(y+x)-4] \}
	    \nonumber   \\
         &&+ \delta''(\hat z) \frac{\lambda_1}{6 m_b^2}~ 
	     [4 \hat q^2 -(y+x)^2]\,\,,
	      \nonumber      \\
\hat W_3 &=& \frac{\delta(\hat z)}{2} 
	        \nonumber   \\
	 &&+ \delta'(\hat z) \{ \frac{5 \lambda_1}{12 m_b^2}~(y+x)
	     + \frac{\lambda_2}{4 m_b^2}~[5(y+x)-12] \}
	      \nonumber    \\
        & &+ \delta''(\hat z) \frac{\lambda_1}{12 m_b^2}~
	     [4 \hat q^2 -(y+x)^2]\,\,,
	      \nonumber   \\
\hat W_4 &=& \delta'(\hat z) \frac{2(\lambda_1+3 \lambda_2)}{3 m_b^2} \,\,,
	     \nonumber   \\
\hat W_5 &=&-\frac{\delta(\hat z)}{2}
	       \nonumber   \\
	 &&-\delta'(\hat z) \{\frac{\lambda_1}{12 m_b^2}~[5(y+x)+8]
	    + \frac{5 \lambda_2}{4 m_b^2}~(y+x) \}
	     \nonumber   \\
         &&- \delta''(\hat z) \frac{\lambda_1}{12 m_b^2}~
	     [4 \hat q^2 -(y+x)^2]\,\,,
	      \nonumber   \\
\hat V_1 &=&-\delta(\hat z)(1-\frac{\lambda_1+3\lambda_2}{4 m_b^2})
		   \nonumber \\
	 & &+\delta'(\hat z) \{\frac{\lambda_1}{6 m_b^2}~[-4-3(y+x)]
	    +\frac{3 \lambda_2}{2 m_b^2}~[2-(y+x)] \}\nonumber \\
	 & &  -\delta''(\hat z)\frac{\lambda_1}{6 m_b^2}~[\hat 4q^2-
	    (y+x)^2]\,\,, \nonumber  \\
\hat V_2 &=&\delta(\hat z)[1-\frac{3(\lambda_1+3\lambda_2)}{4 m_b^2}]
		  \nonumber \\
	 & & +\delta'(\hat z)\{\frac{5 \lambda_1}{6 m_b^2}~(y+x)
	   -\frac{3\lambda_2}{2 m_b^2}~[2-(y+x)]\} \nonumber \\
	 & & +\delta''(\hat z) \frac{\lambda_1}{6 m_b^2}~[4 \hat q^2-
	   (y+x)^2]\,\,, \nonumber \\ 
\hat S   &=&\delta(\hat z)\{[\frac{1}{2}-\frac{1}{4}(y+x)]
	       +\frac{\lambda_1+3\lambda_2}
	  {8 m_b^2}~[4-(y+x)]\} \nonumber \\
	 & &+\delta'(\hat z)[\frac{\lambda_1+3\lambda_2}{4m_b^2}~
	     \{\frac{\hat q^2}{3}+(y+x)-\frac{5}{6}(y+x)^2\}]
		     \nonumber \\
	 & &+\delta''(\hat z)\frac{\lambda_1}{24 m_b^2}~[2-(y+x)]
	   [4 \hat q^2-(y+x)^2]\,\,, 
\ee
where
\begin{equation}
\hat z = 1+\hat q^2-\rho_c-y-x+i\epsilon\,\,.
\end{equation}
\newpage
\appendix
 
The expressions of the lepton energy spectra 
${\cal J}_1$, ${\cal J}_2$ and ${\cal J}_3$ are given by
\be
{\cal J}_1 &=& 2\sqrt{y^2-4\rho_{\tau}} 
	\nonumber \\
 &&   \times \left[~x_0^3~[y^2-3y(1+\rho_{\tau}) +8\rho_{\tau}]
       + x_0^2[-3y^2 + 6y(1+\rho_{\tau})-12 \rho_{\tau}]\right. 
	   \nonumber \\
 && ~~~~-\frac{\lambda_2 x_0}{m_b^2 (1+\rho_{\tau}-y)}
        [ 5x_0^2(y^3 -4 y^2(1+\rho_{\tau})
          +2y(3+7 \rho_{\tau})+4 \rho_{\tau} (\rho_{\tau}-5)) 
	   \nonumber \\
 && ~~~~~~~~+ 3x_0 (-5y^3+y^2(17+15\rho_{\tau})-y(24+46\rho_{\tau})
           -\rho_{\tau}(18 \rho_{\tau}-70)) 
	    \nonumber \\
 && ~~~~~~~~+3(5y^3-10y^2(1+\rho_{\tau})+4y(3+4\rho_{\tau})
        +16\rho_{\tau}(\rho_{\tau}-2)) ]    
	    \nonumber \\
 && ~~~~+\frac{\lambda_1}{3m_b^2 (1+\rho_{\tau}-y)^2}
      [3(y^4-2y^3(1+\rho_{\tau})+8y \rho_{\tau}(1+\rho_{\tau})
        -16{\rho_{\tau}}^2) 
	   \nonumber \\
 && ~~~~~~~~+2x_0^3 (y^4-5y^3(1+\rho_{\tau})+2y^2(5+11\rho_{\tau}
        +5{\rho_{\tau}}^2)-40y \rho_{\tau}(1+\rho_{\tau})
	\nonumber \\
 && ~~~~~~~~-2\rho_{\tau}(5-38\rho_{\tau}+5{\rho_{\tau}}^2)) 
	    \nonumber \\
 && ~~~~~~~~+3x_0^2(-2y^4+8y^3(1+\rho_{\tau})-y^2(15+28\rho_{\tau}
          +15{\rho_{\tau}}^2)+52y\rho_{\tau}(1+\rho_{\tau})
	  \nonumber \\
 && ~~~~~~~~+18\rho_{\tau}(1-6\rho_{\tau}+{\rho_{\tau}}^2)) 
	    \nonumber \\
 && ~~~~~~~~+6x_0(y^4-3y^3(1+\rho_{\tau})+y^2(5+6\rho_{\tau}
          +5{\rho_{\tau}}^2)-12y\rho_{\tau}(1+\rho_{\tau})
	  \nonumber \\
 && ~~~~~~~~\left.-8\rho_{\tau}(1-4\rho_{\tau}+{\rho_{\tau}}^2)) ] \right]~~,
	  \nonumber \\
{\cal J}_2&=&\sqrt{\rho_\tau}\sqrt{y^2-4\rho_\tau}
	\nonumber \\
 && \times \left[~x_0^2~(12+12 \rho_\tau-12y)\right.
         \nonumber \\
 &&~~~~+\frac{\lambda_2}{m_b^2(1+\rho_\tau-y)}[x_0(-36y^2+36y(3+\rho_\tau)
	 -72(1+\rho_\tau)) \nonumber \\
 &&~~~~~~~~~~~~+x_0^2(9y^2-36y+27+18\rho_\tau-9{\rho_\tau}^2)]
	  \nonumber \\
 &&~~~~+\frac{\lambda_1}{m_b^2(1+\rho_\tau-y)^2}[4y^3-4y^2(1+\rho_\tau)
	-16y\rho_\tau + 16\rho_\tau(1+\rho_\tau)
	\nonumber \\
 &&~~~~~~~~+x_0(4y^3-16y^2(1+\rho_\tau)+2y(6+28\rho_\tau+6{\rho_\tau}^2)
      - 32\rho_\tau(1+\rho_\tau))\nonumber \\
 &&~~~~~~~~+x_0^2(y^3-y^2(1+\rho_\tau) +y(3-10\rho_\tau+3{\rho_\tau}^2)
	 \nonumber \\
 &&~~~~~~~~\left.-3+7\rho_\tau+7{\rho_\tau}^2-3{\rho_\tau}^3) ]\right]\,\,,
	 \nonumber \\
{\cal J}_3&=&\frac{1}{2}\sqrt{y^2-4 \rho_\tau}
	       \nonumber \\
 &&\times \left[~x_0^3~(y^2-3y(1+\rho_\tau)+8\rho_\tau)
    -3x_0^2(y^2-3y(1+\rho_\tau)+4\rho_\tau) \right.
      \nonumber \\
 &&~~~~+\frac{\lambda_2}{2m_b^2(1+\rho_\tau-y)}[-6x_0(5y^3-2y^2(3+5\rho_\tau)
     +10y \rho_\tau+4{\rho_\tau}^2 )
     \nonumber \\
 && ~~~~~~~~+3x_0^2(13y^3-y^2(47+39\rho_\tau)+2y(15+64\rho_\tau+3{\rho_\tau}^2)
     -4\rho_\tau(19+3\rho_\tau))
     \nonumber \\
 &&~~~~~~~~-x_0^3(13y^3-52y^2(1+\rho_\tau)+y(69+182\rho_\tau+9{\rho_\tau}^2)
     +16\rho_\tau(14-\rho_\tau))
     \nonumber \\
 &&~~~~+\frac{\lambda_1}{3m_b^2(1+\rho_\tau-y)^2}[~3~(y^4-2y^3(1+\rho_\tau)
     +8y\rho_\tau(1+\rho_\tau)-16{\rho_\tau}^2)
     \nonumber \\
 &&~~~~~~~~+3x_0(2y^4-6y^3(1+\rho_\tau)+2y^2(5+9\rho_\tau+5{\rho_\tau}^2)
     -6y\rho_\tau(5+7\rho_\tau)-4\rho_\tau(4-19\rho_\tau+{\rho_\tau}^2))
     \nonumber \\
 &&~~~~~~~~+3x_0^2(-14y^4+y^3(83+56\rho_\tau)
                    -y^2(123+304\rho_\tau+75{\rho_\tau}^2) 
     \nonumber \\
 && ~~~~~~~~+y(45+358\rho_\tau+361{\rho_\tau}^2+24{\rho_\tau}^3)-6\rho_\tau(
      11+50\rho_\tau+15{\rho_\tau}^2))
      \nonumber \\
 &&~~~~~~~~+2x_0^3(7y^4-35y^3(1+\rho_\tau)+2y^2(26+77\rho_\tau+26{\rho_\tau}^2)
      \nonumber \\
 &&~~~~~~~~\left. -2y(9+95\rho_\tau +95{\rho_\tau}^2+9{\rho_\tau}^3)+2\rho_\tau(19+86\rho_\tau
	     +19{\rho_\tau}^2))] \right]\,\,,
\ee
with
\be
x_0 = 1- \frac{\rho_c}{1+\rho_{\tau}-y}
\ee
\newpage
\appendix

The decay rates $\Gamma_1$, $\Gamma_2$ and 
$\Gamma_3$ are given by
\begin{eqnarray}
\Gamma_1&=& \sqrt{\lambda}
   \left[ (1+\frac{\lambda_1}{2 m_b^2})
     \{1-7(\rho_c+\rho_{\tau}) -7({\rho_c}^2+{\rho_{\tau}}^2)
      +({\rho_c}^3+{\rho_{\tau}}^3) \right.
	 \nonumber \\
    &&~~~~~~~~~~+\rho_c \rho_{\tau}[12-7(\rho_c+\rho_{\tau})] \}
	 \nonumber \\
    &&~~~~+\frac{3 \lambda_2}{2 m_b^2}
        \{-3+5(\rho_c+\rho_{\tau})
         -19({\rho_c}^2+{\rho_{\tau}}^2)
         +5({\rho_c}^3+{\rho_{\tau}}^3)
	 \nonumber \\
   &&~~~~~~~~~~+7\rho_c \rho_{\tau}[4-5(\rho_c+\rho_{\tau})]\} \bigg]
	      \nonumber \\
  && +12(1+\frac{\lambda_1+3\lambda_2}{2 m_b^2})
       [{\rho_c}^2 \mbox{ln}\frac{(1+\rho_c-\rho_{\tau}
        +\sqrt{\lambda})^2}{4 \rho_c}
        +{\rho_{\tau}}^2 \mbox{ln} \frac{(1+\rho_c-\rho_{\tau}
       +\sqrt{\lambda})^2}{4\rho_{\tau}}]
	     \nonumber \\
  && -12(1+\frac{\lambda_1+15\lambda_2}{2 m_b^2})
         {\rho_c}^2{\rho_{\tau}}^2 \mbox{ln}
         \frac{(1-\rho_{\tau}-\rho_c +\sqrt{\lambda})^2}
          {4 \rho_{\tau} \rho_c}\,\,,
	  \nonumber \\
\Gamma_2&=&\sqrt{\rho_\tau}\sqrt{\lambda}
	  \bigg[~2-10 \rho_c + 20 \rho_\tau-4{\rho_c}^2 
	  + 2 {\rho_\tau}^2 -10 \rho_c \rho_\tau 
	  \nonumber \\
        &&~~~~~~+\frac{\lambda_1}{2m_b^2}(-1+5  \rho_c-10 \rho_\tau
	 +2{\rho_c}^2-{\rho_\tau}^2+5  \rho_c \rho_\tau)
	 \nonumber \\
        &&~~~~~~\left.+\frac{3\lambda_2}
			    {2m_b^2}(-9+21 \rho_c-30 \rho_\tau
	-6{\rho_c}^2+3{\rho_\tau}^2-15 \rho_c \rho_\tau) \right]
	\nonumber \\
        &&+3\{{\rho_c}^2[4(1-\rho_\tau)-\frac{\lambda_1}{m_b^2}
	(1-\rho_\tau)-\frac{3\lambda_2}{m_b^2}(1+3 \rho_\tau)\}
	\mbox{ln}\frac{(1+\rho_c-\rho_\tau+\sqrt{\lambda})^2}
	{4 \rho_c} \nonumber \\
       &&+3 \{-8 \rho_\tau+16 \rho_c \rho_\tau-8 {\rho_c}^2\rho_\tau
       -8{\rho_\tau}^2
       +\frac{2\lambda_1}{m_b^2}(\rho_\tau-2 \rho_c \rho_\tau
       +{\rho_c}^2 \rho_\tau+{\rho_\tau}^2)
		\nonumber \\
       &&~~~~+\frac{6\lambda_2}{m_b^2}(5 \rho_\tau-2 \rho_c \rho_\tau
       -3 {\rho_c}^2 \rho_\tau +{\rho_\tau}^2 \}
       \mbox{ln}\frac{(1+\rho_\tau-\rho_c+\sqrt{\lambda})^2}
       {4 \rho_\tau}\,\,,
	     \nonumber \\
\Gamma_3&=&\sqrt{\lambda}
	  \left[~\frac{1}{4}~[1-7(\rho_c+\rho_\tau)-7({\rho_c}^2
	  +{\rho_\tau}^2)+({\rho_c}^3+{\rho_\tau}^3) \right.
	   \nonumber \\
       &&~~~~~~~ +\rho_c \rho_\tau(12-7(\rho_c+\rho_\tau))] 
	  \nonumber \\
       &&~~~~+\frac{3\lambda_1}{8 m_b^2}[39+123 \rho_c-97\rho_\tau
       -9 {\rho_c}^2-37{\rho_\tau}^2+3({\rho_c}^3+{\rho_\tau}^3)
	 \nonumber \\
       &&~~~~~~~-\rho_c \rho_\tau(16+21(\rho_c+\rho_\tau))]
	 \nonumber \\
       &&~~~~+\frac{3\lambda_2}{4m_b^2}[47+95 \rho_c-61 \rho_\tau
	-13 {\rho_c}^2 -{\rho_\tau}^2+3({\rho_c}^3+{\rho_\tau}^3)
	  \nonumber \\
       &&~~~~~~~+\rho_c \rho_\tau (32-21(\rho_c+\rho_\tau))] \bigg]
	\nonumber \\
      &&+\{3{\rho_c}^2(1-{\rho_\tau}^2)
	\nonumber \\
      &&~~~-\frac{3\lambda_1}{2m_b^2}(2+26\rho_c-4\rho_\tau+11{\rho_c}^2
       +2{\rho_\tau}^2-36\rho_c\rho_\tau+14\rho_c{\rho_\tau}^2
       +9{\rho_c}^2{\rho_\tau}^2)
	\nonumber \\
      &&~~~+\frac{6\lambda_2}{m_b^2}(-1-8\rho_c+2\rho_\tau-2{\rho_c}^2
	-{\rho_\tau}^2+4 \rho_c\rho_\tau+2\rho_c{\rho_\tau}^2
	-3{\rho_c}^2{\rho_\tau}^2)\}
	\nonumber \\
     &&~~\times \mbox{ln}\frac{(1+\rho_c-\rho_\tau+\sqrt{\lambda})^2}
			      {4\rho_c}
	\nonumber \\
     &&+\{ 6 {\rho_\tau}^2(1-{\rho_c}^2)
     +\frac{3\lambda_1}{2m_b^2}{\rho_\tau}^2(46-28\rho_c-18{\rho_c}^2)
       \nonumber \\
    &&~~~+\frac{18\lambda_2}{m_b^2}{\rho_\tau}^2
	    (1+2\rho_c-3{\rho_c}^2)\}
    \mbox{ln}\frac{(1+\rho_\tau-\rho_c+\sqrt{\lambda})^2}{4\rho_\tau}
\ee
with 
\begin{equation}
\lambda = 1 - 2(\rho_c+\rho_{\tau})+(\rho_c-\rho_{\tau})^2 \,\,.
\end{equation}

\newpage
\begin{figure}
\caption{
Relevant diagrams to obtain amplitudes 
$T^{\mu \nu}$, $T^{\mu}$, $T$. (a), (b) correspond to W exchange terms,
(c) - (f) the interference terms and (g), (h) charged Higgs
exchange terms.
}
\label{figone}
\end{figure}


\begin{references}
\bibitem{HH} {\it For review}: J. F. Gunion, H. E. Haber and S. Dawson,
{\sl The Higgs Hunter's Guide}, (Addison-Wessley, London, 1990).
\bibitem{CLEO} M. Battle {\it et al}., CLEO Collaboration, in
{\it Proceedings of the Joint International Lepton-Photon Symposium
and Europhysics Conference of High-Energy Physics}, Geneva, 1991,
edited by S. Hegarty, K. Potter and E. Quercigh (World Scientific, Singapore,
1992).
\bibitem{park} G. T. Park, Phys. Rev. {\bf D 50}, 599 (1994). 
\bibitem{putzer} A. Putzer, {\it talk delivered at the 5th International 
Symposium on Heavy Flavour Physics}, Canada, 1993, HD--IHEP/93--03.
\bibitem{l3} M. Acciarri {\it et al.}, L3 Collaboration, Phys. Lett. {\bf B 332}, 201 (1994).
\bibitem{chay} J. Chay, H. Georgi and B. Grinstein, Phys. Lett. {\bf B 247}, 399 (1990).
\bibitem{bigi} I. I. Bigi, M. Shifman, N. G. Uraltsev and A. Vainshtein, 
Phys. Rev. Lett. {\bf 71}, 496 (1993).
\bibitem{manohar} A. V. Manohar and M. B. Wise, Phys. Rev. {\bf D 49}, 1310 (1994). 
\bibitem{blok} B. Blok, L. Koyrakh and M. Shifman, Phys. Rev. {\bf D 49}, 3356 (1994). 
\bibitem{falk} A. F. Falk, Z. Ligeti, M. Neubert and Y. Nir, Phys. Lett. {\bf B 326}, 145 (1994).
\bibitem{balk} S. Balk, J. G. K$\ddot{\mbox o}$rner, D. Prijol and K. Schilcher, MZ--TH/93--32.
\bibitem{koyrakh} L. Koyrakh, Phys. Rev. {\bf D 49}, 3379 (1994). 
\bibitem{pdg} K. Hikasa {\it et al.},Particle Data Group, Phys. Rev. {\bf D 45}, S1 (1992). 
\bibitem{isidori} G. Isidori, Phys. Lett. {\bf B 298}, 409 (1993).
\bibitem{aleph} D. Buskulic {\it et al.}, ALEPH Collaboration, Phys. Lett. {\bf B 298}, 479 (1993).
\end{references}
\end{document}